\documentclass[pre]{revtex4}
\usepackage{graphicx}
\begin{document}
\title{Self-Organized Criticality of Domain Walls and Magnetization Curve}
\author{Hidetsugu Sakaguchi and Yue Zhao}
\affiliation{Department of Applied Science for Electronics and Materials,
Interdisciplinary Graduate School of Engineering Sciences, Kyushu
University, Kasuga, Fukuoka 816-8580, Japan}
\begin{abstract}
We propose a kind of Ginzburg--Landau equation with quenched randomness. There is a pinning--depinning transition in the system when the external magnetic force is changed. The transition is self-organized when the external magnetic field is slowly changed under the demagnetizing effect. The total magnetization increases stepwise and the probability distribution of the increase in the total magnetization approximately obeys a power law. A hysteresis loop is obtained when the external magnetic field is changed reciprocally. In our model, the coercivity in the magnetization curve is expressed as the critical value for the pinning-depinning transition.
\end{abstract}
\maketitle
\section{Intoduction}
Self-organized criticality has been intensively studied since the proposal of the sand pile model by Bak et al.~\cite{Bak}. It suggests that critical states naturally appear in slowly driven dissipative systems, which might explain the power laws widely observed in nature. 
 On the other hand, there is a pinning-depinning transition in random growth problems. A typical example is random interface growth described by the quenched Kardar--Parisi--Zhang equation~\cite{Leschhorn}. We showed that a critical state of the pinning-depinning transition is naturally obtained when the external force is weakened as the interface grows~\cite{Sakaguchi1}. The self-organized criticality of the pinning-depinning transition was applied to a random block-spring model of earthquakes, and a power law of the slip size of faults was naturally observed~\cite{Sakaguchi2}. 

In this paper, we apply the self-organized criticality to magnetic systems. 
A ferromagnet is microscopically composed of many magnetic domains.  There is  a domain wall between two magnetic domains of different orientations.   
Magnetic domain walls can be trapped in some nonmagnetic impurities. The trapped domain walls move intermittently if the external magnetic field is increased. The intermittent motion of domain walls causes Barkhausen noises. These noises are audible as sound noises when the magnetic materials are connected to a speaker. It was observed that a histogram of the jump size of the Barkhausen noises obeys a power law with an exponent of around 1.5~\cite{Bertotti}.  The Barkhausen noises are interpreted as an example of the self-organized criticality by some authors~\cite{Cote}. The relationship between the Barkhausen noise and the depinning transitions of domain walls was studied using interface equations by several authors~\cite{Urbach, Zapperi}. In this paper, we propose a Ginzburg--Landau-type model and discuss the domain wall motion, the Barkhausen noise, and the hysteresis loop of magnetization.  In general, the magnetization curve is rather complicated,  since various factors such as domain nucleation, spin rotation, magnetocrystalline anisotropy, and domain wall motion are involved. For example, there are usually many magnetic domains and the complicated interaction among many domain walls should be considered. Furthermore, it is known that the effect of the rotation of the magnetization vector is important near the saturation magnetization at a strong magnetic field. In this paper, however, we consider only the effect of the motion of a single domain wall in the case that there is only one axis of easy magnetization. The effects of multidomain walls and the rotation of the magnetization direction are neglected for the sake of simplicity. 
\section{Model Equation}
We use the Ginzburg--Landau equation in two dimensions, because it is a simple model equation and widely studied for phase transitions. 
The model equation is expressed as
\begin{equation}
\frac{\partial m}{\partial t}=m-m^3+D\nabla^2 m+\epsilon H,
\end{equation}
where $m$ is the local magnetization, $D$ denotes the coupling strength required to make the local magnetization uniform, $H$ is the external magnetic field, and $\epsilon$ is a parameter that determines the relationship between the magnetic field and magnetization. The total magnetization is defined as $M=\int\int m(x,y)dxdy$.  
For $H=0$, there are two stable uniform solutions $m=\pm 1$, and there is a solution of $m(x,y)={\rm tanh}\{\kappa(y-y_0)\}$, where $\kappa=1/\sqrt{2D}$. The solutions $m=\pm 1$ represent uniform states with inverse magnetization directions.  The solution $m(x,y)={\rm tanh}\{\kappa(y-y_0)\}$ represents a domain wall between the two domains of $m=\pm 1$. The parameter $D$ determines the width of the domain wall. If $H$ is positive (negative), the domain wall moves as the domain of $m=1$ ($m=-1$) becomes dominant, and a spatially uniform state of $m=1$ ($m=-1$) is finally obtained. 

In ferromagnets, many magnetic domains are created so as to reduce the magnetostatic energy. The magnetostatic energy is the self-energy due to the interaction with the magnetic field created by the magnetization. This magnetic field generated by the magnetization reduces the magnetization and the effect is called the demagnetizing effect. For a uniformly magnetized ellipsoidal object of magnetization $I$, the internal magnetic field $H$ is expressed as $H=H_0-\gamma(I/\mu_0)$, where $H_0$ is the external uniform magnetic field, $\mu_0$ is the vacuum permeability, and $\gamma$ is called the demagnetizing factor. For example, $\gamma=1/3$ for a uniformly magnetized sphere. 

The motion of a magnetic domain wall is disturbed by various impurities and grain boundaries of the crystal. There are several model equations including random magnetic fields; however, in this paper, we consider the effect of nonmagnetic impurities where $m$ tends to be zero. By taking these effects into consideration, we propose a model equation of the form
\begin{equation}
\frac{\partial m}{\partial t}=m-m^3+D\nabla^2 m+\epsilon(H-\beta \bar{m})-\alpha(x,y)m,
\end{equation}
where $\beta$ denotes the demagnetizing coefficient, $\bar{m}=M/S$ with total area $S$, and $\alpha(x,y)$ takes a  random value larger than 1 at nonmagnetic impurities.  Since $1-\alpha(x,y)<0$,  $m$ tends to become zero at the position of a nonmagnetic impurity if $D=\epsilon=0$. 
The numerical simulations are performed on a rectangular lattice of $L_x\times L_y$ with lattice constant 1, $\nabla^2m$ is replaced with $m(i+1,j)+m(i-1,j)+m(i,j+1)+m(i,j-1)-4m(i,j)$, and $S=L_x\times L_y$. Periodic boundary conditions are imposed at $i=1$ and $L_x$, and no-flux boundary conditions are imposed at $j=1$ and $L_y$ to set a single horizontal domain wall as an initial condition.    

First, we show the pinning phenomenon of the domain wall in this model equation. The impurity is set within a small circle in a system of $L_x\times L_y=200\times 200$, that is, $\alpha(i,j)=1.25$ for $(i-L_x/2)^2+(j-L_y/2-10)^2\le 10$. The initial domain wall is set to $j=L_y/2-10$, and the parameters are $\epsilon=1$, $\beta=1$, and $H$ is increased stepwise as $H=-0.1+0.001 n$ ($n=1,2,\cdots$). Figure 1(a) shows the positions of the domain wall at $H=-0.1+0.02n^{\prime}$ ($n^{\prime}=1,2,\cdots)$. The impurity region is marked with rhombi near $(i,j)=(100,110)$. The domain wall moves upward with $H$, but it is pinned to the impurity region for $0.026<H<0.375$, and then depinning occurs for $H>0.376$. Figure 1(b) shows the $y$ coordinate of the domain wall at $i=L_x/2$ as a function of $H$. The nearly flat region of $0.026<H<0.375$ in Fig.~1(b) represents the pinning phenomenon.

\begin{figure}[h]
\begin{center}
\includegraphics[height=4.5cm]{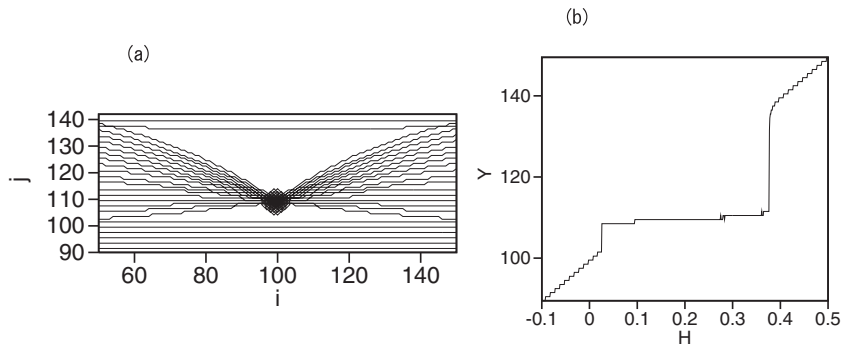}
\end{center}
\caption{(a) Domain walls for $H=-0.1+0.02n^{\prime}$ ($n^{\prime}=1,2,\cdots$). (b) $y$-coordinate of the domain wall at $i=L_x/2$ as a function of $H$.}
\label{fig1}
\end{figure}
\section{Self-Organized Criticality of Domain Wall Motion}
We shows the pinning-depinning transition of quenched random systems where $\alpha(i,j)$ takes a random number between 1 and $1.2$ with probability $p$ and 0 with probability $1-p$ for $\beta=0$.  If the demagnetizing effect is absent, there is a pinning-depinning transition as $H$ is increased. That is, the domain wall is pinned to the randomness for $H<H_c$, and the domain wall moves with a nonzero average velocity for $H>H_c$. 
We have calculated the growth velocity of the magnetization as a function of $H$ for $\epsilon=1$. Here, the average velocity is evaluated at $v_M=\langle (1/L_x)dM/dt\rangle$. Here, the average $\langle \cdots\rangle$ implies both the 
long-time average and the average with respect to four ensembles of different randomness. When the domain wall is pinned, $v_M$ is zero, and $v_M$ takes a nonzero value when the domain wall moves with a nonzero average velocity. 
Figure 2 shows the relationship between $H$ and $v_M$ for (a) $p=0.05$ and $D=0.5$, and (b) $p=0.2$ and $D=0.25$.  Continuous transitions are observed. The pinning-depinning transition occurs at $H=H_c\sim 0.024$ for $p=0.05$ and $D=0.5$ and at $H=H_c\sim 0.084$ for $p=0.2$ and $D=0.25$.  
\begin{figure}[h]
\begin{center}
\includegraphics[height=4.5cm]{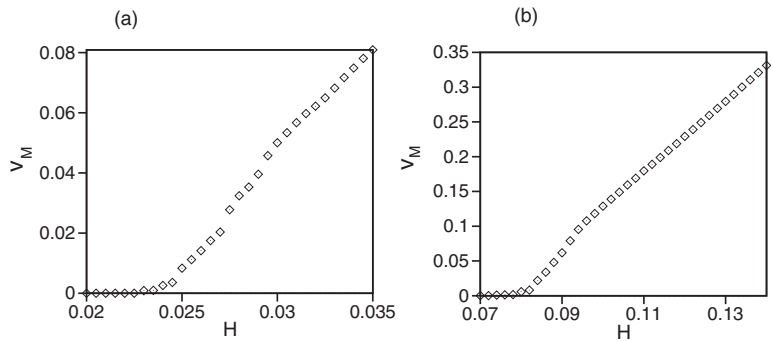}
\end{center}
\caption{Pinning-depinning transition for (a) $p=0.05$ and $D=0.5$ and (b) $p=0.2$ and $D=0.25$.}
\label{fig2}
\end{figure}

If the demagnetizing effect is taken into consideration and $H$ is slowly increased, the system tends to stay around the pinning-depinning transition, and the critical state for the pinning-depinning transition is approximately realized. The mechanism of the self-organization of the critical state is qualitatively explained as follows, although there is no theoretical proof. 
 If $\epsilon(H-\beta \bar{m})$ is larger than the critical value, the domain wall moves forward. If the domain wall moves forward for a certain constant $H$, $\bar{m}$ increases and $\epsilon(H-\beta \bar{m})$ decreases until $\epsilon(H-\beta \bar{m})$ becomes below the critical value. On the other hand, if $\epsilon(H-\beta \bar{m})$ is smaller than the critical value, the domain wall is pinned and $\bar{m}$ does not increase. Since $H$ is slowly increased under a constant $\bar{m}$, $\epsilon(H-\beta \bar{m})$ increases slowly and exceeds the critical value, and then, the domain wall begins to move forward. In any case, $\epsilon(H-\beta \bar{m})$ fluctuates around the critical value.   

Figure 3 shows numerical results for $p=0.05$, $\beta=1$, $\epsilon=1$, and $D=0.5$. The system size is $L_x\times L_y=1000\times 500$. $H$ is increased stepwise as $H=-0.75+0.0001n$ ($n=1,2,\cdots,1500)$. Long-time numerical simulation of Eq.~(2) is performed for each value of $H$ until the stationary state is obtained.  Figure 3(a) shows several snapshots of the domain wall for $0<H<0.05$.  It is seen that the domain wall is pinned by the nonmagnetic impurities for $0<x<300$. Figure 3(b) shows time evolutions of $H-\bar{m}$. For $H>-0.7$, $H-\bar{m}$ is fluctuating around 0.022, which is close to the critical value 0.024. Figure 3(c) shows time evolutions of the change $\Delta M=M(n+1)-M(n)$. Magnetic avalanches of various sizes occur. Figure 3(d) shows the size distribution of $\Delta M$. The dashed line denotes the power law of $1/(\Delta M)^{1.14}$. Five random ensembles are used to calculate the size distribution.   
\begin{figure}[h]
\begin{center}
\includegraphics[height=4.cm]{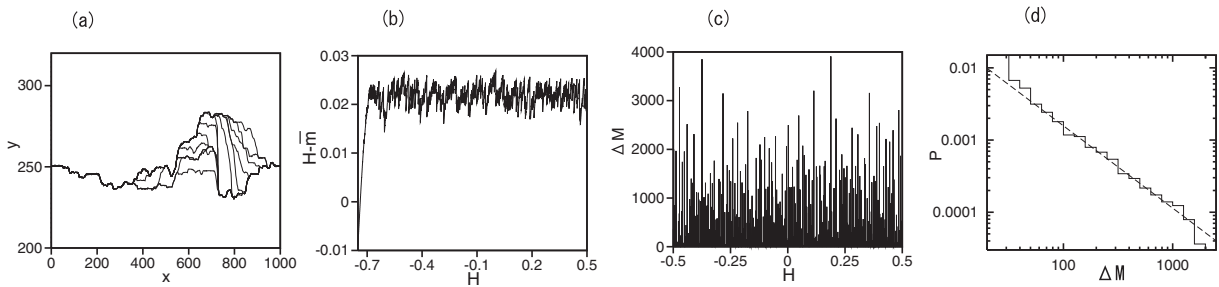}
\end{center}
\caption{(a) Several snapshots of domain wall for $p=0.05$ and $D=0.5$. (b) Time evolutions of $H-\bar{m}$. (c) Time evolutions of $\Delta M$. (d) Size distribution $P(\Delta M)$.}
\label{fig3}
\end{figure}

Figure 4 shows numerical results for $p=0.2$, $\beta=1$, $\epsilon=1$, and $D=0.25$. Figure 4(a) shows time evolutions of $H-\bar{m}$ for $H=-0.75+0.0001n$ ($n=1,2,\cdots,1500)$. For $H>-0.5$, $H-\bar{m}$ is fluctuating around 0.083, which is close to the critical value 0.084. Figure 4(b) shows time evolutions of the change $\Delta M=M(n+1)-M(n)$. Magnetic avalanches of various sizes occur. Figure 4(c) shows the size distribution of $\Delta M$ calculated using five random samples. The dashed line denotes the power law of $1/(\Delta M)^{1.14}$. 

Several authors studied similar probability distributions for the quenched Edwards--Wilkinson equation (qEW equation) and Kardar--Paris--Zhang equation with quenched randomness (KPZQ equation), and evaluated the exponent to be $1.08$ for the qEW equation and 1.25 for the KPZQ equation. The domain wall in our model equation can move in any direction. Our model might be close to the KPZQ equation, since the effect of the obliquely moving interface is somewhat incorporated in the nonlinear term of the KPZQ equation~\cite{Sakaguchi2,Lacombe,Chen}. However, the exponent of the size distribution takes a value between the exponents of the qEW equation and the KPZQ equation.  
\begin{figure}[h]
\begin{center}
\includegraphics[height=4.cm]{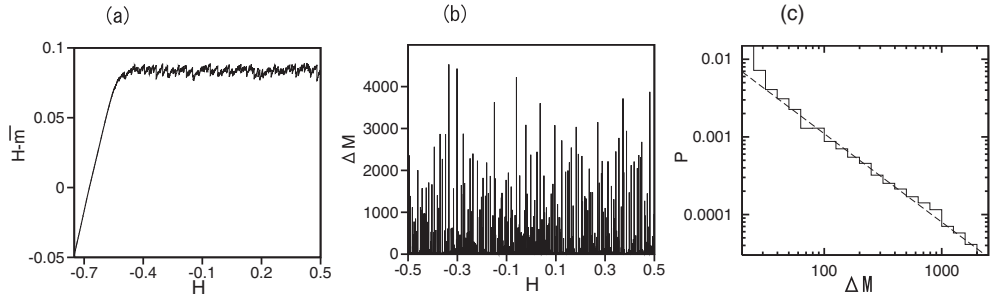}
\end{center}
\caption{(a) Time evolutions of $H-\bar{m}$ for $p=0.2$ and $D=0.25$. (b) Time evolutions of $\Delta M$. (c) Size distribution $P(\Delta M)$.}
\label{fig4}
\end{figure}

\begin{figure}[h]
\begin{center}
\includegraphics[height=4.cm]{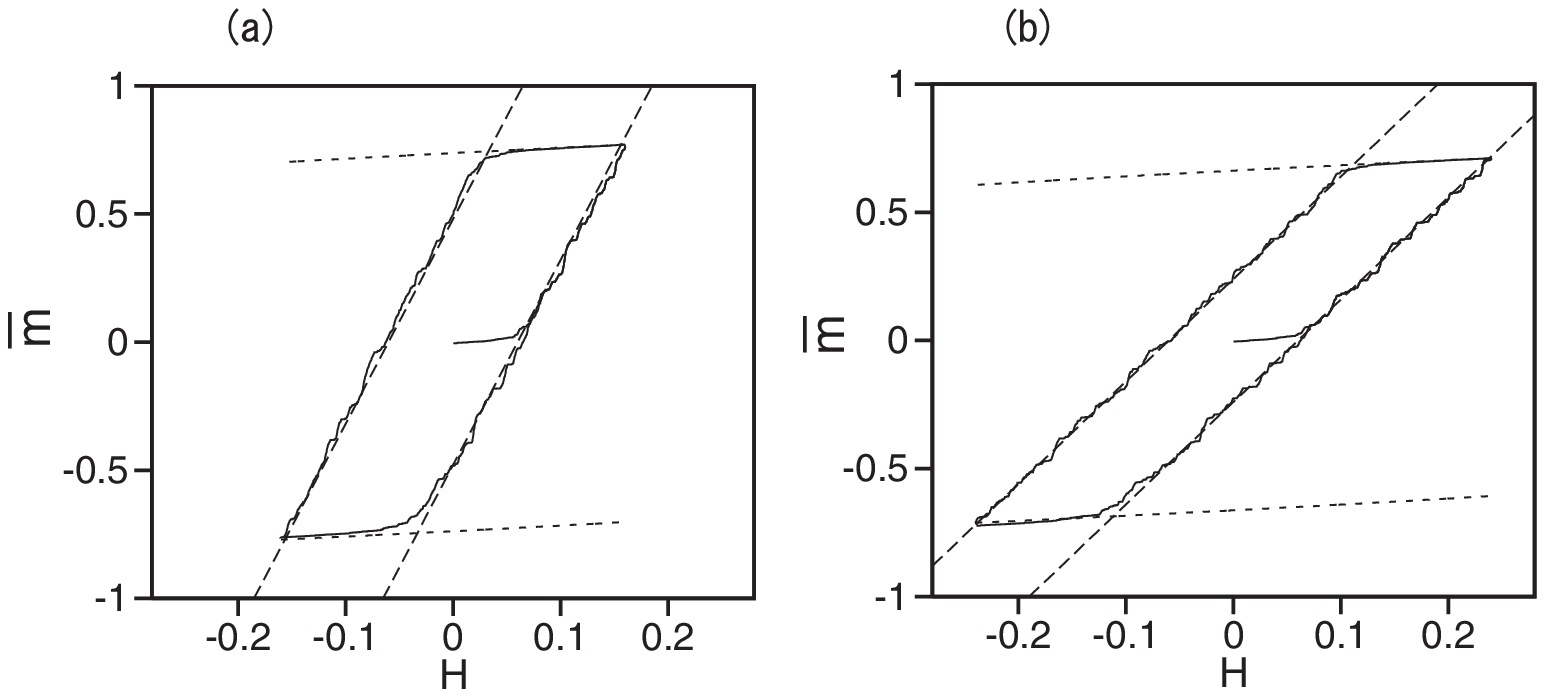}
\end{center}
\caption{Magnetization curves by changing $H$ at (a) $\beta=0.125$ and $N=800$  and (b) $\beta=0.25$ and $N=1200$  for $p=0.05$, $\epsilon=0.4$, and $D=0.5$. The dashed curves are $\bar{m}=(H\pm 0.024/\epsilon)/\beta$. The dotted curves are $\bar{m}=r_{+}\{(1-p)m_{+}+pm_{+}^{\prime}\}+(1-r_{+})\{(1-p)m_{-}+pm_{-}^{\prime}\}$ using the solutions to Eqs.~(3).}
\label{fig5}
\end{figure}
\begin{figure}[h]
\begin{center}
\includegraphics[height=4.cm]{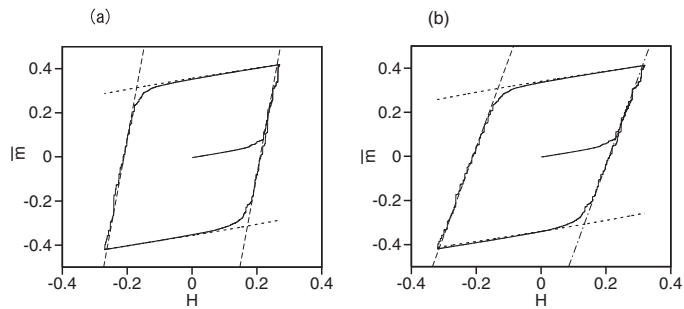}
\end{center}
\caption{Magnetization curves obtained by changing $H$ at (a) $\beta=0.125$ and $N=1350$  and (b) $\beta=0.25$ and $N=1600$ for $p=0.2$, $\epsilon=0.4$, and $D=0.25$. The dashed curves are $\bar{m}=(H\pm 0.084/\epsilon)/\beta$. The dotted curves are $\bar{m}=r_{+}\{(1-p)m_{+}+pm_{+}^{\prime}\}+(1-r_{+})\{(1-p)m_{-}+pm_{-}^{\prime}\}$ using the solutions to Eqs.~(3).}
\label{fig6}
\end{figure}
\section{Magnetization Curve}
The magnetization curve can be calculated by changing $H$ as $H=0.0002\times n$ for $1\le n\le N$, $H=0.0002\times (2N-n)$ for $N< n\le 3N$, and $H=0.0002\times (n-4N)$ for $3N< n\le 5N$.  The system size is $L_x\times L_y=500\times 500$. Initially, a domain wall is set at $j=L_y/2$. Even for the maximum $H$ of $H=0.0002N$, the domain wall does not reach $j=L_y$. That is, the saturation magnetization is not obtained even for the maximum value of $H$. Figures 5(a) and 5(b)  show  magnetization curves $\bar{m}(H)$ at (a) $\beta=0.125$  and (b) $0.25$ for $p=0.05$, $\epsilon=0.4$, and $D=0.5$.
  The magnetization curves consist of four curves. These curves correspond to the upward motion of the domain wall, an almost pinned state, the downward motion of the domain wall, and another almost pinned state. 
In the two jagged curves with a relatively large slope, the domain walls move upward or downward. The jagged curves are approximated with $\bar{m}=(H\pm 0.024/\epsilon)/\beta$ (dashed lines), which implies that the self-organized criticality is approximately satisfied because $\epsilon(H-\beta\bar{m})\simeq H_c\sim \pm 0.024$. 
In the two smooth and nearly horizontal curves, the domain wall is almost pinned. The ratio of the $m>0$ domain is around $r_{+}=0.888$ at $\beta=0.125$ and $r_{+}=0.858$ at $\beta=0.25$. A kind of mean-field approximation can be applied to approximate the relationship between $H$ and $\bar{m}$ if the ratios $r_{+}$ and $r_{-}=1-r_{+}$ are fixed along the nearly horizontal curves. In the $m>0$ domain, the local magnetization is assumed to take a constant value $m_{+}$ at magnetic sites and $m_{+}^{\prime}$ at nonmagnetic impurity sites. Similarly, in the $m<0$ domain, the local magnetization is assumed to take a constant value $m_{-}$ at magnetic sites and $m_{-}^{\prime}$ at nonmagnetic impurity sites. 
The local magnetizations $m_{\pm}$ and $m_{\pm}^{\prime}$ approximately satisfy  
\begin{eqnarray}
m_{+}-m_{+}^3+\epsilon(H-\beta \bar{m})+4D\{(1-p)m_{+}+pm_{+}^{\prime}-m_{+}\}&=&0,\nonumber\\
m_{-}-m_{-}^3+\epsilon(H-\beta \bar{m})+4D\{(1-p)m_{-}+pm_{-}^{\prime}-m_{-}\}&=&0,\nonumber\\
-0.1\cdot m_{+}^{\prime}-m_{+}^{\prime 3}+\epsilon(H-\beta \bar{m})+4D\{(1-p)m_{+}+pm_{+}^{\prime}-m_{+}^{\prime}\}&=&0,\nonumber\\
-0.1\cdot m_{-}^{\prime}-m_{-}^{\prime 3}+\epsilon(H-\beta \bar{m})+4D\{(1-p)m_{-}+pm_{-}^{\prime}-m_{-}^{\prime}\}&=&0.
\end{eqnarray}
Here, the following two assumptions are used: the average value of $1-\alpha(x,y)$ at nonmagnetic impurities is equal to $-0.1$, and the average value of $m$ is expressed as $\bar{m}=r_{+}\{(1-p)m_{+}+pm_{+}^{\prime}\}+(1-r_{+})\{(1-p)m_{-}+pm_{-}^{\prime}\}$. The dotted nearly horizontal curves in Figs.~5(a) and 5(b) denote  $\bar{m}$ calculated from numerical solutions to Eq.~(3) for (a) $\beta=0.125$ and (b) $\beta=0.25$. Fairly good agreement with direct numerical simulations is observed.  
As $H$ is decreased with $H=0.0002\times(2N-n)$ from the maximum value $H=0.0002\times N$ along the upper nearly horizontal curve, $H$ reaches a critical point satisfying $\epsilon(H-\beta\bar{m})\sim -0.024$. At that time, the domain wall begins to move downward, and then the magnetization curve changes from the upper nearly horizontal curve to the left jagged curve.  The coercivity is an important quantity for a permanent magnet because it is closely related to its 
strength. In our model, the coercivity is expressed as $H_c/\epsilon$ using the critical value for the pinning-depinning transition as seen in Fig.~5(a).
      
Similar behaviors are observed for $p=0.2$, $\epsilon=0.4$, and $D=0.25$.  Figures 6(a) and 6(b) show two magnetization curves $\bar{m}(H)$ at (a) $\beta=0.125$ and (b) $0.25$.  The magnetization curves also consist of two jagged curves with a relatively large slope and smooth curves with a small slope. The jagged curves are approximated as $\bar{m}=(H\pm 0.084/\epsilon)/\beta$ (dashed curves), which implies self-organized criticality. In the two smooth curves of the smaller slope, the domain wall is almost pinned. The ratio of the $m>0$ domain is around $r_{+}=0.706$ at $\beta=0.125$ and $r_{+}=0.703$ at $\beta=0.25$. The dotted curves with a smaller slope in Figs.~6(a) and 6(b) denote  theoretical curves of $\bar{m}$ calculated from numerical solutions to Eq.~(3) for (a) $\beta=0.125$ and (b) $\beta=0.25$. Fairly good agreement is seen for this parameter set. 
 The coercivity is approximated by $H_c/\epsilon=0.084/\epsilon$. The coercivity is larger for this parameter set than in the case shown in Fig.~5 since the pinning effect by quenched randomness is larger.  
\section{Summary}  
We have proposed a Ginzburg--Landau-type model equation with quenched randomness. There is a pinning-depinning transition in the system when the external magnetic force is changed. When the external magnetic field is slowly changed under the demagnetizing effect, the total magnetization increases stepwise and intermittently, which induces  Barkhausen noises. The probability distribution of the increase in total magnetization approximately obeys a power law with an exponent of around 1.14. A hysteresis loop corresponding to the magnetization curve is obtained when the external magnetic field is changed reciprocally. In our model, the coercivity in the magnetization curve is expressed as the critical value for the pinning-depinning transition, and two jagged curves with stepwise increments are approximately expressed using the condition of self-organized criticality. The magnetization curves where the domain wall is almost pinned to the magnetic impurities have been evaluated using a mean-field approximation.

In contrast to previous model studies of Barkhausen noises, we have performed numerical simulations of the simple Ginzburg-Landau equation and shown that the Barkhausen noises can be interpreted as the self-organized criticality to the depinning transition. In this paper, we focused on the motion of a single domain wall in nonmagnetic impurities to clarify the relationship to the self-organized criticality. In realistic permanent magnets, there are many domain walls, and the creation and annihilation of various magnetic domains have to be considered. Furthermore, the rotation of the direction of the magnetization is another important factor to understand the magnetization curves more quantitatively.  The simple Ginzburg--Landau equation can be generalized to models including vector variables, and the problems including  multidomain walls can also be studied using the generalized Ginzburg-Landau models. In the future, we would like to study such more complicated cases using the generalized models.      

\end{document}